\documentclass[reprint,unsortedaddress,superscriptaddress,footinbib,amsmath,amssymb,aps,pra,showkeys,showpacs]{revtex4-2}
\usepackage{graphicx}
\usepackage{xcolor}
\usepackage[colorlinks=true, linkcolor=blue, citecolor=blue, urlcolor=blue]{hyperref}
\usepackage{amsmath}
\usepackage{amssymb}
\usepackage{braket}
\usepackage{bm}

\usepackage{mathrsfs}

\newcommand{\qda}{\text{QD}_{1}}
\newcommand{\qdb}{\text{QD}_{2}}

\newcommand{\Neg}{\mathcal{N}}
\newcommand{\taup}{\tau_{\mathrm{p}}} 

\begin{document}

\title{Dissipation as a Resource for Entanglement via High-Spin Mediator Purification}

\author{E. M. Fernandes}
\email{Contact author: ellen.fernandes@ufu.br}
\affiliation{Instituto de F\'{i}sica, Universidade Federal de Uberl\^andia,
38400-902, MG, Brazil}
\author{F. M. Souza}
\email{Contact author: fmsouza@ufu.br}
\affiliation{Instituto de F\'{i}sica, Universidade Federal de Uberl\^andia,
38400-902, MG, Brazil}

\date{\today}

\begin{abstract}
We study entanglement generation between two spin qubits mediated by a single, thermally mixed, high-spin ($S=5/2$) magnetic impurity, using the Lindblad master equation to compare coherent and dissipative control of the mediator. 
We show that no coherent drive, however strong or resonant, can purify the mediator beyond its thermal value --- a direct consequence of unitary evolution exactly preserving state purity --- while a dissipative channel purifies it up to the limit set by the optical pumping cycle itself, $\mathrm{Tr}[\rho_M^2]\approx0.8$. 
Modeling this channel with rates fixed by the measured optical pumping of a single Mn spin, we find that the entanglement it enables is not monotonic in the pump strength: it peaks at a finite rate and then decreases, since the same channel leaks decoherence to the qubits through the coupling that mediates their entanglement. 
At the experimental pumping rate, a single $80$~ns pulse purifies the mediator, the resulting polarization persists for a timescale three orders of magnitude longer than the entangling dynamics, and the first-recurrence negativity rises from $0.18$ to $0.64$ at $12$~K, remaining above $0.6$ up to $30$~K.
These results establish dissipation, not as noise to be suppressed, but as the very resource that makes robust entanglement generation possible.
\end{abstract}

\maketitle

\section{Introduction}
\label{sec:intro}

Quantum information processing exploits coherent superposition and entanglement to enable computational and communication tasks with no classical counterpart. Realizing this potential in a physical platform requires qubits that can be initialized, coherently controlled, and entangled on demand, while remaining sufficiently isolated from their environment to preserve quantum coherence. Entanglement, in particular, is not merely a byproduct of quantum dynamics but the central resource underlying quantum computation, quantum teleportation, and quantum key distribution~\cite{horodecki2009entanglement}, which places its controlled generation and protection against decoherence at the heart of the field.

Among the many physical platforms explored for quantum information processing, semiconductor quantum dots have emerged as a leading candidate, owing to their compatibility with mature nanofabrication technology and the long coherence times achievable by encoding qubits in individual electron or hole spins~\cite{lossdivincenzo1998quantum}. Since the original proposal of spin qubits in coupled quantum dots, the field has matured into a broad experimental and theoretical effort spanning electrical initialization, coherent single- and two-qubit control, and high-fidelity readout~\cite{hanson2007spins,burkard2023semiconductor}, establishing semiconductor spin qubits as one of the most versatile solid-state architectures for quantum information processing.

A recurring strategy for entangling qubits that are not directly coupled is to introduce an auxiliary quantum degree of freedom, or mediator, through which an effective interaction between the qubits is generated.
One realization of this idea is the \textit{spin bus}, in which a finite spin chain with strong, permanent exchange interactions connects external qubits through local couplings~\cite{friesen2007spinbus}. 
Another approach uses collective magnetic excitations of a ferromagnet, which can couple spatially separated spins through dipolar interactions~\cite{trifunovic2013ferromagnet}.
Although distinct, these proposals rely on collective degrees of freedom or multiple ancillary spins to establish interactions between the qubits.

In contrast, this work considers a single quantum object with its own discrete spin degree of freedom as a mediator of entanglement between spatially separated qubits. 
Beyond extending connectivity, such a mediator offers the possibility of independently manipulating its state, providing an additional means of controlling the dynamics of the system. 
A physical realization is a CdTe quantum dot containing a single Mn$^{2+}$ ion, whose spin state can be optically prepared and read out through the spectral structure arising from its exchange interaction with an exciton~\cite{besombes2004probing,besombes2006optical,legall2010initialization}. Optical initialization of the Mn spin has been demonstrated even at zero magnetic field under resonant excitation, with efficiencies of up to $75\%$ on timescales of tens of nanoseconds~\cite{legall2010initialization}. 
An applied field of about $1$~T suppresses the fast zero-field relaxation of the Mn spin, caused by hyperfine mixing of its eigenstates, extending its storage time to hundreds of microseconds in the dark~\cite{goryca2009optical}. 
The optical control of individual atom-like spins in semiconductors~\cite{awschalom2013spintronics}, and of single magnetic impurities in particular, defines the emerging field of \textit{solotronics}~\cite{kobak2014designing}. Although these systems have been studied mainly as spin memories and probes of local exciton fine structure~\cite{goryca2009optical,kobak2014designing,besombes2006optical}, their use as active mediators of entanglement between spatially separated qubits remains comparatively unexplored. 
%The optical addressability of the Mn ion therefore provides a means of controlling the mediator independently of the qubits it connects.

In this work, we investigate entanglement generation between two spatially separated spin qubits mediated by a single high-spin magnetic impurity, with particular attention to the role of thermal mixing of the mediator's spin state --- an unavoidable feature of any realistic, finite-temperature implementation.
We show that coherent control applied to the mediator alone cannot eliminate this thermal mixing, since unitary evolution preserves its state purity; genuine dissipation is required instead.
This places the present work within a broader line of research in which dissipation, rather than being treated purely as a source of decoherence to be suppressed, is engineered as an active resource for state preparation and entanglement generation~\cite{verstraete2009,krauter2011}.
We investigate this approach through an effective dissipative channel associated with an optical pumping cycle, with rates obtained from a microscopic model of the pumping cycle with experimental parameters~\cite{smolenski2018mechanisms}, including reverse processes that limit the attainable spin polarization.
We further show that this resource is subject to its own trade-off: because the dissipative channel that purifies the mediator necessarily couples to it through the same interaction that mediates the qubits' entanglement, pushing it too hard becomes counter-productive, and an optimal pump strength emerges naturally from the dynamics. 
The remainder of this paper is organized as follows: Sec.~\ref{sec:theory} presents the theoretical formalism, including a rigorous justification of why purification requires a dissipative rather than coherent channel; Sec.~\ref{sec:results} reports our numerical results; and Sec.~\ref{sec:conclusions} summarizes our conclusions.

\section{Theoretical formalism}
\label{sec:theory}

\subsection{Model Hamiltonian}
\label{sec:theory-hamiltonian}

\begin{figure}
\centering
\includegraphics[width=1.0\columnwidth]{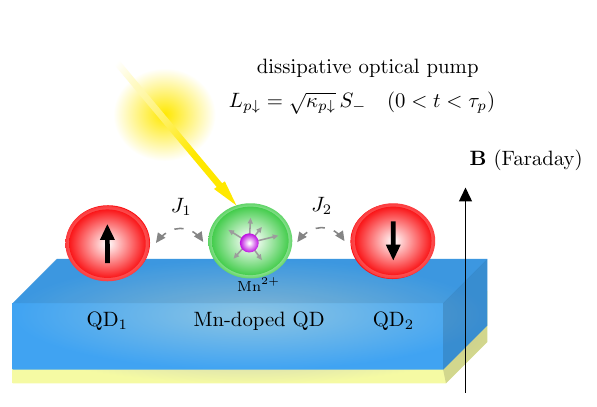}
\caption{Schematic of the physical system. Two spin-$1/2$ qubits ($\qda$, $\qdb$, red) are exchange-coupled, with strengths $J_1$ and $J_2$ [Eq.~\eqref{eq:V}], to a common high-spin ($S=5/2$) Mn$^{2+}$ mediator confined within its own quantum dot (green, outlined to distinguish it from the bare CdTe host), the three dots individually embedded within a common CdTe host. The qubits are initialized in a definite product spin state (solid arrows), while the mediator starts in a thermally mixed state with no preferred orientation (faint arrow fan). A uniform external field \textbf{B}, applied in Faraday configuration (along the growth axis), sets both Zeeman splittings $\varepsilon_{\mathrm{QD}}$ and $\varepsilon_M$ [Eq.~\eqref{eq:zeeman}]. 
During a finite window $0<t<\taup$, an additional dissipative channel, $L_{\mathrm p\downarrow}=\sqrt{\kappa_{\mathrm p\downarrow}}\,S_-$ with a weaker backward counterpart $L_{\mathrm p\uparrow}=\sqrt{\kappa_{\mathrm p\uparrow}}\,S_+$ [Eq.~\eqref{eq:pump}], acts on the mediator alone, without directly perturbing the qubits, physically realized through a quasi-resonant optical pumping cycle via the confined exciton, at $570$--$600$~nm (yellow beam) \cite{goryca2009optical}. (Color online)}
\label{fig:system}
\end{figure}

Figure~\ref{fig:system} sketches the physical system considered throughout this work. We consider two spin-$1/2$ qubits, $\qda$ and $\qdb$, each described by Pauli operators $\bm{\sigma}_i=(\sigma_{ix},\sigma_{iy},\sigma_{iz})$ ($i=1,2$), exchange-coupled to a common high-spin magnetic impurity (the mediator) of spin $S$, described by spin operators $\mathbf{S}=(S_x,S_y,S_z)$ acting on a $(2S+1)$-dimensional Hilbert space.

The total Hamiltonian, defined on $\mathcal{H}_{\qda}\otimes\mathcal{H}_{\qdb}\otimes\mathcal{H}_\text{M}$, is
\begin{equation}
H = H_0 + V,
\label{eq:Htot}
\end{equation}
with
\begin{equation}
H_0 = \varepsilon_{\mathrm{QD}}\left(\sigma_{1z}+\sigma_{2z}\right) + \varepsilon_\text{M} S_z,
\label{eq:H0}
\end{equation}
\begin{equation}
V = -J_1\,\mathbf{s}_1\!\cdot\!\mathbf{S} - J_2\,\mathbf{s}_2\!\cdot\!\mathbf{S},
\qquad \mathbf{s}_i\equiv\bm{\sigma}_i/2.
\label{eq:V}
\end{equation}
Both Zeeman-like splittings in Eq.~\eqref{eq:H0} arise from the same uniform external field, applied in Faraday configuration, but with independent $g$-factors for the two subsystems,
\begin{equation}
\varepsilon_{\mathrm{QD}} = g_{\mathrm{dot}}\,\mu_\text{B} B, \qquad \varepsilon_M = g_{\mathrm{Mn}}\,\mu_\text{B} B,
\label{eq:zeeman}
\end{equation}
with $\mu_B$ the Bohr magneton. We take $g_{\mathrm{Mn}}=2.0$, the standard value for a substitutional Mn$^{2+}$ ion, essentially material-independent since it originates from a localized, atomic-like $d$-electron spin state~\cite{smolenski2015cdse}, and $g_{\mathrm{dot}}=1.5$, appropriate for an electron confined in a II-VI quantum dot~\cite{besombes2004probing}. Since $|g_{\mathrm{dot}}|<g_{\mathrm{Mn}}$, a single field $B$ splits the mediator's sublevels more strongly than the qubits', without requiring either an independently tuned coupling or a spatially inhomogeneous field. $J_1$ and $J_2$ are the exchange coupling strengths between each qubit and the mediator (the microscopic sp-d exchange interaction), which we treat as fixed, always-active physical couplings, never switched off externally. Throughout this work we set $\hbar=1$ (see Appendix~\ref{app:units} for how real-time units are recovered), so energies and rates share the same units, and we focus on $S=5/2$, appropriate for a substitutional Mn$^{2+}$ ion, unless noted otherwise.

\subsection{Initial state and thermal mixing}
\label{sec:theory-thermal}

The qubits are initialized in the factorized logical state $\rho_Q(0)=\ket{0}\!\bra{0}\otimes\ket{1}\!\bra{1}$. We assume this state is actively prepared prior to the protocol, a standard requirement for quantum information processing, using techniques demonstrated for single electron spins in quantum dots, such as spin-selective tunneling to a reservoir~\cite{elzerman2004single} or optical pumping of the electron spin~\cite{atature2006quantum}. 
Alternatively, the qubits could be initialized electrically, through gate-controlled, on-demand single-electron injection from the leads into each dot, as recently proposed for charge qubits~\cite{fernandes2025pulsed}; with spin-polarized (e.g., ferromagnetic) leads, the same scheme would load each dot with an electron of definite spin. 
The mediator, in contrast, is an auxiliary element that is not prepared by any such step: having equilibrated with its local environment prior to the entangling protocol, it is initialized in the thermal (Gibbs) state at temperature $T$,
%The qubits are initialized in the factorized logical state $\rho_Q(0)=\ket{0}\!\bra{0}\otimes\ket{1}\!\bra{1}$. The mediator, having equilibrated with its local environment prior to the entangling protocol, is initialized instead in the thermal (Gibbs) state at temperature $T$,
\begin{equation}
\rho_M^{\mathrm{th}}(T) = \frac{e^{-H_M/k_BT}}{Z}, \qquad
H_M = \varepsilon_M S_z,
\label{eq:rhoMth}
\end{equation}
\begin{equation}
Z = \mathrm{Tr}\!\left[e^{-H_M/k_BT}\right],
\label{eq:partitionfn}
\end{equation}
with $k_B$ Boltzmann's constant. The full initial state of the composite system is the product $\rho(0)=\rho_Q(0)\otimes\rho_M^{\mathrm{th}}(T)$. Since $\rho_M^{\mathrm{th}}(T)$ is diagonal in the $S_z$ basis with population spread over all $2S+1$ sublevels except in the $T\to0$ limit, this thermal mixing is the central obstacle addressed in this work: the exchange process that entangles the two qubits via Eq.~\eqref{eq:V} is most effective when the mediator occupies its extremal sublevel $m=-S$, a condition satisfied only at very low temperature or, as we show below, when actively prepared.

\subsection{Open-system dynamics}
\label{sec:theory-lindblad}

The coupling of the composite system to its environment is modeled through a Lindblad (GKSL) master equation for the full density matrix $\rho(t)$,
\begin{equation}
\frac{d\rho}{dt} = -i[H,\rho] + \sum_k \mathcal{D}[L_k]\rho,
\label{eq:lindblad}
\end{equation}
\begin{equation}
\mathcal{D}[L]\rho \equiv L\rho L^\dagger - \tfrac{1}{2}\{L^\dagger L,\rho\},
\label{eq:dissipator}
\end{equation}
solved numerically using QuTiP~\cite{johansson2012qutip,johansson2013qutip2}. Two groups of dissipative channels enter Eq.~\eqref{eq:lindblad}.
%Population relaxation of the qubits' own electron spin (as opposed to pure dephasing) is neglected throughout, since the corresponding relaxation time is much longer than the nanosecond timescales of the entangling and purification dynamics considered here: for a single Mn atom in a CdTe quantum dot, electron spin-relaxation times reaching the millisecond range at cryogenic temperature have been calculated theoretically, three orders of magnitude longer than the room-temperature values~\cite{petrovic2012spin}.
% A comparação com a temperatura ambiente não serve ao argumento, porque o que importa é a comparação com a escala de nanossegundos da dinâmica, por isso ele foi tirada daqui e não foi re-adicionada no tópico da termalizacao

\textit{Always-on environment.}--- Independently of the purification protocol below, the following channels remain active throughout the entire evolution, representing the mediator's intrinsic relaxation toward its own thermal equilibrium and the pure dephasing unavoidably affecting both the mediator and the qubits:
\begin{equation}
\begin{aligned}
L_{\Gamma_\downarrow}&=\sqrt{\Gamma_\downarrow}\,S_-, &\qquad L_{\Gamma_\uparrow}&=\sqrt{\Gamma_\uparrow}\,S_+, \\
L_{\gamma_M}&=\sqrt{\gamma_M}\,S_z, &\qquad L_{\gamma_Q}^{(i)}&=\sqrt{\gamma_Q}\,\sigma_{iz},
\end{aligned}
\label{eq:cops-baseline}
\end{equation}
with $i=1,2$ and $S_\pm=S_x\pm iS_y$ the mediator's spin ladder operators. The rates $\Gamma_\downarrow$ and $\Gamma_\uparrow$ satisfy detailed balance,
\begin{equation}
\Gamma_\uparrow/\Gamma_\downarrow = e^{-\varepsilon_M/k_BT},
\label{eq:detailedbalance}
\end{equation}
which guarantees that, absent the exchange coupling $V$ and the purification pulse introduced below, the mediator relaxes toward the same thermal state $\rho_M^{\mathrm{th}}(T)$ used to initialize it [Eq.~\eqref{eq:rhoMth}]. The rates $\gamma_M$ and $\gamma_Q$ are independent, phenomenological pure-dephasing rates for the mediator and for each qubit, respectively. Because Eq.~\eqref{eq:lindblad} describes purely Markovian (exponential) decay, $\gamma_Q$ is more naturally compared to a spin-echo $T_2$ than to the free-induction $T_2^*$, which is instead dominated by quasi-static (non-Markovian) hyperfine noise: in the original Loss--DiVincenzo GaAs implementation, $T_2^*\sim10\,$ns was measured for free induction, but $T_2\gtrsim1\,\mu$s was recovered once this quasi-static contribution was refocused by a spin-echo pulse~\cite{petta2005coherent}. Our choice of $\gamma_Q$ matches this echo-limited value closely.

\textit{Purification pulse.}--- During a finite window $0<t<\taup$, we supplement the always-on environment of Eq.~\eqref{eq:cops-baseline} with two additional dissipative channels acting on the mediator alone,
\begin{equation}
L_{\mathrm{p}\downarrow} = \sqrt{\kappa_{\mathrm p\downarrow}}\,S_-, \qquad L_{\mathrm{p}\uparrow} = \sqrt{\kappa_{\mathrm p\uparrow}}\,S_+ ,
\label{eq:pump}
\end{equation}
which pump the mediator preferentially toward its extremal sublevel $m=-S$, on a timescale $\sim1/\kappa_{\mathrm p\downarrow}$. This pulse represents the quasi-resonant optical pumping demonstrated for a single Mn spin in a quantum dot~\cite{legall2009optical,goryca2009optical}: photocreated excitons exchange spin with the Mn ion and carry angular momentum away upon recombination. The backward channel $L_{\mathrm p\uparrow}$ accounts for the processes within the same cycle that push the Mn spin the opposite way, chiefly the hole--Mn flip-flop enabled by heavy-hole/light-hole mixing~\cite{smolenski2018mechanisms}; it limits the attainable polarization, so that the steady state under pumping is $p_{m+1}/p_m=\kappa_{\mathrm p\uparrow}/\kappa_{\mathrm p\downarrow}$ rather than a pure state. Both rates were obtained by fitting Eq.~\eqref{eq:pump} to a microscopic model of the pumping cycle of a single Mn$^{2+}$ ion in a CdTe quantum dot under quasi-resonant $\sigma^+$ excitation, with exchange constants, carrier relaxation and recombination times, and spin-flip probabilities taken from Ref.~\cite{smolenski2018mechanisms} (not shown). The fit gives $\kappa_{\mathrm p\downarrow}=0.013\,\mu$eV, corresponding to $\hbar/\kappa_{\mathrm p\downarrow}\approx50\,$ns, consistent with the $20$--$100\,$ns orientation times measured by Goryca \textit{et al.}~\cite{goryca2009optical}, and $\kappa_{\mathrm p\uparrow}/\kappa_{\mathrm p\downarrow}=0.104$, which limits the mediator's purity under pumping to $\mathrm{Tr}[\rho_M^2]\approx0.81$ (population $0.89$ in $m=-S$). Crucially, the exchange coupling $V$ [Eq.~\eqref{eq:V}] is never switched off during the pulse, consistent with it being a fixed physical interaction rather than an externally controllable parameter; the qubit-mediator entangling dynamics and the mediator's purification therefore proceed concurrently rather than in separate stages. Sec.~\ref{sec:results-optimal} examines how the entanglement depends on the pump strength, scanning $\kappa_{\mathrm p\downarrow}$ at fixed ratio $\kappa_{\mathrm p\uparrow}/\kappa_{\mathrm p\downarrow}$.

\subsection{Purification requires genuine dissipation}
\label{sec:theory-theorem}

Before turning to the numerical results, we note a structural fact that motivates the specific form of Eq.~\eqref{eq:pump}. For any unitary $U$, conjugation $\rho\to U\rho U^\dagger$ leaves the eigenvalues of $\rho$, and hence its purity $\mathrm{Tr}[\rho^2]$, exactly invariant, since $U$ only rotates the eigenbasis without altering the spectrum. Consequently, \emph{no} Hamiltonian acting on the mediator alone, however strong or elaborately time-dependent, can increase $\mathrm{Tr}[\rho_M^2]$ beyond its initial thermal value: purifying a mixed state fundamentally requires a genuine loss channel, one that carries entropy away to an external reservoir. This is why the purification mechanism must enter Eq.~\eqref{eq:lindblad} as Lindblad dissipators, Eq.~\eqref{eq:pump}, rather than as an additional term in $H$; a microscopic derivation showing precisely how this dissipator emerges from coupling the mediator to a continuum of electromagnetic modes --- in contrast to a single mode, which cannot purify it --- is given in Appendix~\ref{app:jc}. We verify this distinction numerically in Sec.~\ref{sec:results-theorem}.

\subsection{Entanglement and purity quantifiers}
\label{sec:theory-quantifiers}

We quantify the mediator's degree of purification at any instant $t$ through the purity of its reduced density matrix,
\begin{equation}
\mathrm{Tr}[\rho_M(t)^2], \qquad \rho_M(t) = \mathrm{Tr}_{\qda,\qdb}[\rho(t)],
\label{eq:purity}
\end{equation}
which ranges from $1/(2S+1)$ (maximally mixed) to $1$ (pure state).

To quantify the entanglement generated between the two qubits, we use the negativity, computed from the two-qubit reduced density matrix $\rho_Q(t)=\mathrm{Tr}_M[\rho(t)]$:
\begin{equation}
\Neg(t) = \big\| \rho_Q(t)^{T_1} \big\|_1 - 1 = 2\sum_i \left|\lambda_i^{-}\right|,
\label{eq:negativity}
\end{equation}
where $\rho_Q(t)^{T_1}$ denotes the partial transpose of $\rho_Q(t)$ with respect to $\qda$, $\|X\|_1=\mathrm{Tr}[\sqrt{X^\dagger X}]$ is the trace norm, and $\lambda_i^{-}$ are the negative eigenvalues of $\rho_Q(t)^{T_1}$. For a pair of qubits this normalization gives $\Neg\in[0,1]$, with $\Neg=0$ for separable states and $\Neg=1$ for maximally entangled (Bell) states; being based on the positive partial transpose criterion, it constitutes a full entanglement monotone for two-qubit states.

\begin{table*}
\caption{Default parameter values used throughout Sec.~\ref{sec:results} (natural units, $\hbar=1$). Both Zeeman splittings follow from a common Faraday field $B=1\,$T [Eq.~\eqref{eq:zeeman}]. The third column gives the real-time equivalent of each energy/rate; the last column lists comparable experimental values for a single Mn spin in a quantum dot, where available.}
\begin{ruledtabular}
\begin{tabular}{llcl}
Parameter & Value & Real-time equivalent & Comparable experimental value \\
\hline
Faraday field, $B$ & $1\,$T & --- & standard laboratory field \\
Mediator $g$-factor, $g_{\mathrm{Mn}}$ & $2.0$ & --- & Mn$^{2+}$ in a II-VI quantum dot~\cite{smolenski2015cdse,cosacchi2022deterministic} \\
Qubit $g$-factor, $g_{\mathrm{dot}}$ & $1.5$ & --- & electron in a II-VI quantum dot~\cite{besombes2004probing,cosacchi2022deterministic} \\
Mediator Zeeman splitting, $\varepsilon_M$ & $115.8\,\mu$eV & $0.0357\,\mathrm{ns}$ period ($28.0\,\mathrm{GHz}$) & --- \\
QDs Zeeman splitting, $\varepsilon_{\mathrm{QD}}$ & $86.8\,\mu$eV & $0.0476\,\mathrm{ns}$ period ($21.0\,\mathrm{GHz}$) & --- \\
Exchange couplings, $J_1=J_2$ & $1\,\mu$eV & $4.14\,\mathrm{ns}$ period ($0.242\,\mathrm{GHz}$) & --- \\
Boltzmann constant, $k_B$ & $86.17\,\mu$eV/K & --- & --- \\
Mediator relaxation rate, $\Gamma_\downarrow$ & $1.65\times10^{-6}\,\mu$eV & $400\,\mu\mathrm{s}$ lifetime & directly measured at $B=1\,$T~\cite{goryca2009optical} \\
Mediator dephasing rate, $\gamma_M$ & $5\times10^{-4}\,\mu$eV & $1.32\,\mu\mathrm{s}$ lifetime & ---\footnotemark[1] \\
Qubit dephasing rate (each), $\gamma_Q$ & $5\times10^{-4}\,\mu$eV & $1.32\,\mu\mathrm{s}$ lifetime & $T_2\gtrsim1\,\mu\mathrm{s}$ (spin echo, GaAs)~\cite{petta2005coherent} \\
Pump rate, $\kappa_{\mathrm p\downarrow}$ & $0.013\,\mu$eV & $51\,\mathrm{ns}$ lifetime & $20$--$100\,\mathrm{ns}$ optical pumping~\cite{goryca2009optical} \\
Backward pump rate, $\kappa_{\mathrm p\uparrow}$ & $0.104\,\kappa_{\mathrm p\downarrow}$ & $0.49\,\mu\mathrm{s}$ lifetime & pumping model, parameters of~\cite{smolenski2018mechanisms} \\
Pulse duration, $\taup$ & --- & $80\,\mathrm{ns}$ & tens of ns~\cite{legall2009optical,goryca2009optical} \\
\end{tabular}
\end{ruledtabular}
\footnotetext[1]{Population-based optical measurements (e.g., photoluminescence intensity decay, as used to determine $\Gamma_\downarrow$ above) track only the mediator's level populations; to our knowledge, no measurement isolating the Mn ion's own pure-dephasing rate, as distinct from its population relaxation, has been reported.}
\label{tab:parameters}
\end{table*}

\section{Results}
\label{sec:results}

\subsection{Static-field polarization is insufficient}
\label{sec:results-problem}

\begin{figure}
\centering
\includegraphics[width=1.0\columnwidth]{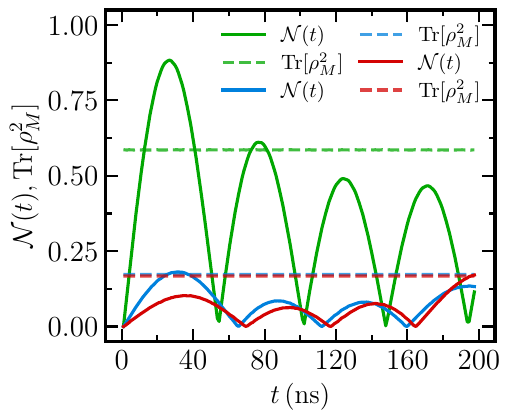}
\caption{Time evolution of the negativity $\Neg(t)$ (solid lines) between $\qda$ and $\qdb$, for $T=1\,\mathrm{K}$~(green), $12\,\mathrm{K}$~(blue), and $30\,\mathrm{K}$~(red), with the mediator ($S=5/2$) permanently subject to the static Faraday field $B=1\,$T and dephasing/relaxation always active, but no purification pulse applied. Dashed lines show the corresponding mediator purity $\mathrm{Tr}[\rho_M^2]$, pinned at its thermal equilibrium value throughout, since the mediator starts already at that fixed point. (Color online)}
\label{fig:problem}
\end{figure}

We begin by characterizing the extent to which the entanglement-generation mechanism identified in Sec.~\ref{sec:theory} survives once realistic decoherence is included, under the external field $B=1\,$T that sets both $\varepsilon_M$ and $\varepsilon_{\mathrm{QD}}$ throughout. Figure~\ref{fig:problem} shows the time evolution of the negativity $\Neg(t)$ between $\qda$ and $\qdb$ for three representative bath temperatures ($T=1$, $12$, and $30\,\mathrm{K}$), with the always-on dissipative environment of Sec.~\ref{sec:theory} active throughout and no purification pulse applied.

Although the field alone already breaks the degeneracy of the mediator's six spin sublevels, its effect on the \emph{equilibrium} population is modest at these temperatures: the thermal purity is $\mathrm{Tr}[\rho_M^2]\approx0.59$ at $T=1\,$K, falling to $\approx0.17$ at both $12$ and $30\,$K, a value barely above the maximally mixed limit of $1/6\approx0.167$. Since the mediator is initialized in, and remains pinned at, this equilibrium value (dashed lines), the resulting negativity degrades correspondingly. We denote by $\Neg_1$ the height of the first recurrence, defined as the first local maximum of $\Neg(t)$ reached shortly after $t=0$, which we adopt throughout this work as the figure of merit for the entangling dynamics' initial strength: it falls from $\approx0.88$ at $T=1\,$K to $\approx0.18$ and $\approx0.10$ at $12$ and $30\,$K. This confirms that a static field of readily achievable laboratory magnitude, applied uniformly and left permanently on, is on its own insufficient to protect the mediated entanglement against thermal mixing across this temperature range; a qualitatively different mechanism is required.

\subsection{Purification requires genuine dissipation: numerical verification}
\label{sec:results-theorem}

\begin{figure}
\centering
\includegraphics[width=1.0\columnwidth]{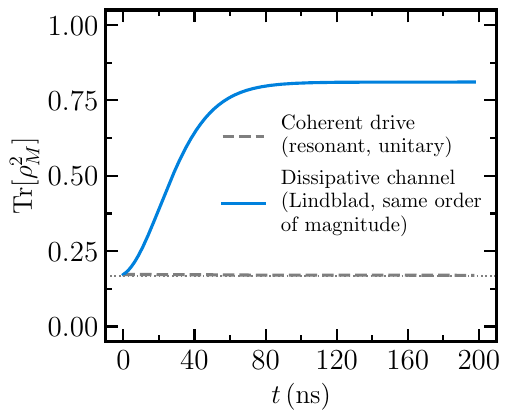}
\caption{Direct comparison of coherent versus dissipative driving of the mediator, both at $T=12\,\mathrm{K}$ and of the same magnitude ($\Omega=\kappa_{\mathrm p\downarrow}=0.013\,\mu$eV). A resonant coherent drive, $H_{\mathrm{drive}}(t)=\Omega(S_+e^{-i\omega t}+S_-e^{+i\omega t})$ with $\omega=\varepsilon_M$ --- the unitary limit of a single-mode (cavity-like) coupling --- leaves the mediator's purity $\mathrm{Tr}[\rho_M^2]$ pinned at its thermal value (dashed line; dotted line marks $1/6$) for the entire duration shown, regardless of the drive being exactly on resonance. The dissipative channels of Eq.~\eqref{eq:pump} (solid line) instead purify the mediator up to the limit $\mathrm{Tr}[\rho_M^2]\approx0.81$ set by $\kappa_{\mathrm p\uparrow}/\kappa_{\mathrm p\downarrow}$, within $\sim100\,$ns.}
\label{fig:theorem}
\end{figure}

The failure of a static field to purify the mediator, established above, raises a sharper question: could \emph{any} coherent drive, however strong and however precisely tuned, achieve what the static field could not? Section~\ref{sec:theory-theorem} answers this rigorously in the negative. We verify this explicitly in Fig.~\ref{fig:theorem}, at $T=12\,$K. We replace the dissipative pump with the coherent, resonant single-mode drive of Eq.~\eqref{eq:jc-single} ($\omega=\varepsilon_M$; see Appendix~\ref{app:jc}), and compare against the dissipative channel of Eq.~\eqref{eq:pump}, using numerically comparable parameters ($\Omega=\kappa_{\mathrm p\downarrow}=0.013\,\mu$eV).

The coherent drive leaves $\mathrm{Tr}[\rho_M^2]$ pinned at its thermal value, $0.173$, throughout the window shown, as it must for any drive strength by the argument of Sec.~\ref{sec:theory-theorem}. The dissipative channels, in contrast, raise the purity from $0.17$ to $0.81$ within $\sim100\,$ns; the plateau lies below unity because the backward channel $L_{\mathrm p\uparrow}$ competes with the pumping, as in the experiments. The distinction between the two is not one of coupling strength or of ``classical'' versus ``quantum'' fields --- both are equally quantum mechanical --- but purely of the \emph{dimensionality of the reservoir}: a single mode always returns its excitation to the mediator (Rabi revivals, bounded Poincar\'e recurrence time), while a continuum carries it away irreversibly (Appendix~\ref{app:jc}). It is this irreversibility, not the strength of the coupling, that enables purification.

\subsection{Optimal pump strength}
\label{sec:results-optimal}

\begin{figure}
\centering 
\includegraphics[width=1.0\columnwidth]{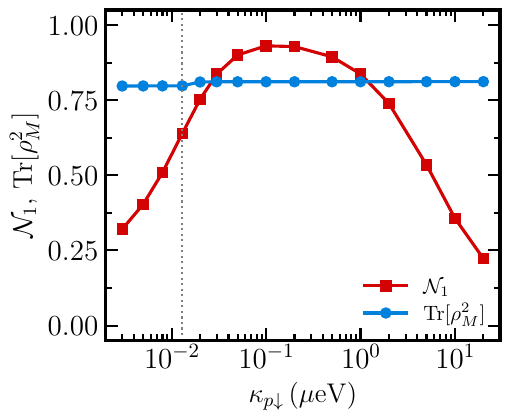}
\caption{Mediator purity $\mathrm{Tr}[\rho_M^2]$ at the end of the pulse (blue circles) and first-recurrence negativity $\Neg_1$ (red squares) as a function of the pump rate $\kappa_{\mathrm p\downarrow}$, at $T=12\,\mathrm{K}$ and fixed $\kappa_{\mathrm p\uparrow}/\kappa_{\mathrm p\downarrow}=0.104$. The dotted line marks the experimental pumping rate, $\kappa_{\mathrm p\downarrow}=0.013\,\mu$eV. The purity stays at $\approx0.80$--$0.81$, the limit set by the backward channel. $\Neg_1$ rises from $0.32$ at $\kappa_{\mathrm p\downarrow}=0.003\,\mu$eV to $0.64$ at the experimental rate and peaks at $\approx0.93$ near $\kappa_{\mathrm p\downarrow}\approx0.1$--$0.2\,\mu$eV, beyond which leaked decoherence reduces it to $0.53$ at $5\,\mu$eV and $\approx0.22$ at $20\,\mu$eV.}
\label{fig:optimal}
\end{figure}

Having established that a genuine dissipative channel is necessary for purification, a natural question follows: is a stronger channel always preferable? Figure~\ref{fig:optimal} shows that it is not. We scan $\kappa_{\mathrm p\downarrow}$ over more than three decades, keeping $\kappa_{\mathrm p\uparrow}/\kappa_{\mathrm p\downarrow}=0.104$ fixed, at $T=12\,$K. The pulse lasts $\taup=80\,$ns at the experimental rate and is lengthened in proportion to $1/\kappa_{\mathrm p\downarrow}$ for slower pumping, so that in every case it covers the first recurrence and saturates the purity.

The purity is essentially the same across the entire range, $0.80$--$0.81$: it is fixed by the ratio $\kappa_{\mathrm p\uparrow}/\kappa_{\mathrm p\downarrow}$, not by the pump speed. The first-recurrence negativity, however, is far from flat. On the low-$\kappa_{\mathrm p\downarrow}$ side it grows with the pump rate, because the first recurrence occurs at $t\approx33\,$ns, while the pulse is still on: a faster pump purifies the mediator further before the qubits reach their first entanglement maximum. It peaks at $\Neg_1\approx0.93$ near $\kappa_{\mathrm p\downarrow}\approx0.1$--$0.2\,\mu$eV and then falls steadily, reaching $\Neg_1\approx0.53$ at $\kappa_{\mathrm p\downarrow}=5\,\mu$eV and $\approx0.22$ at $20\,\mu$eV.

Because the purity is the same across this range, the decline at strong pumping cannot be attributed to incomplete purification but is instead a direct signature of the mechanism identified in Sec.~\ref{sec:theory-theorem}: the dissipators of Eq.~\eqref{eq:pump}, precisely because they are genuine Lindblad channels acting on the mediator, unavoidably leak decoherence into the qubits through the same exchange coupling $V$ that mediates their entanglement. At weak pumping this cost is negligible compared to the benefit of purification; at strong pumping it dominates, degrading the very entanglement the pulse is meant to protect. Purifying the mediator is therefore necessary but not sufficient, and can become actively counterproductive if pushed too far: there is an optimal pump strength, not a monotonic ``stronger is better'' scaling.

The experimental pumping rate lies on the rising side of this curve, about one order of magnitude below the optimum. It already raises $\Neg_1$ from $0.18$ without pumping to $0.64$, and Fig.~\ref{fig:optimal} shows how much more a faster pumping scheme would gain, at fixed $\kappa_{\mathrm p\uparrow}/\kappa_{\mathrm p\downarrow}$. We adopt the experimental value, $\kappa_{\mathrm p\downarrow}=0.013\,\mu$eV, throughout the remainder of this work.

\subsection{Mechanism: a single pulse suffices}
\label{sec:results-mechanism}

At the experimental rate, the pulse duration follows from two time scales. The first recurrence of the negativity occurs at $t\approx33\,$ns, while the pulse is on; scanning $\taup$ shows that $\Neg_1$ grows with the pulse duration up to $\taup\approx40\,$ns and stays at $0.64$ for any longer pulse. A $40\,$ns pulse, however, ends halfway through the purification, leaving the mediator at $\mathrm{Tr}[\rho_M^2]\approx0.64$ for all later times. We therefore adopt $\taup=80\,$ns, which brings the purity to $0.80$, within $2\%$ of its saturated value, and covers the first recurrence; longer pulses add no further purification while continuing to expose the qubits to decoherence leaking through $V$.

\begin{figure}
\centering
\includegraphics[width=1.0\columnwidth]{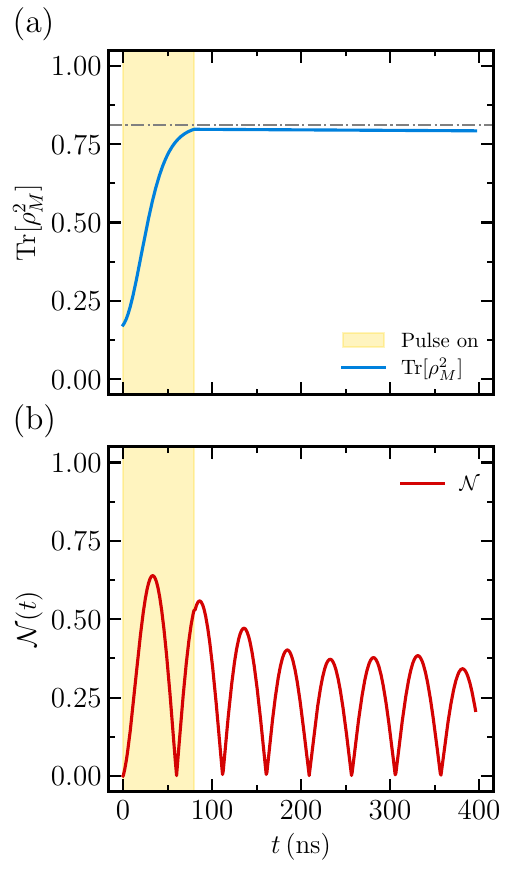}
\caption{Purification-pulse mechanism at $T=12\,\mathrm{K}$, $\kappa_{\mathrm p\downarrow}=0.013\,\mu$eV, $\kappa_{\mathrm p\uparrow}=0.104\,\kappa_{\mathrm p\downarrow}$, with a single pulse of duration $\taup=80\,$ns (shaded region). 
Panel (a): mediator purity $\mathrm{Tr}[\rho_M^2]$, rising from its thermal value ($\approx0.17$) to $0.80$ within the pulse and remaining at $0.79$--$0.80$ for the full $\sim400\,$ns window shown, with no additional pumping; the dash-dotted line marks the saturated value $0.81$.
Panel (b): corresponding negativity $\Neg(t)$. The first recurrence, $\Neg_1=0.64$, occurs during the pulse; the following peaks are $0.56$, $0.47$, $0.40$, and then $0.34$--$0.38$. (Color online)}
\label{fig:mechanism}
\end{figure}

Figure~\ref{fig:mechanism} shows the resulting dynamics for a single pulse applied once at $T=12\,$K. In panel (a) the mediator purity rises during the pulse to $\mathrm{Tr}[\rho_M^2]=0.80$ and then remains at that level for the entire $\sim400\,$ns window shown, without further pumping. This near-permanence is a direct consequence of the mediator's intrinsic relaxation rate $\Gamma_\downarrow$, corresponding to a lifetime of $400~\mu$s (Table~\ref{tab:parameters}), which is more than three orders of magnitude longer than the timescales of the entangling dynamics considered here. Once purified, the mediator therefore has, for all practical purposes, no mechanism left to relax back toward its thermal state within the timescales relevant to the qubits' entangling dynamics.

The negativity, shown in panel (b), reaches its first maximum, $\Neg_1=0.64$, at $t\approx33\,$ns, while the mediator is still being purified: the dissipative channel and the entangling dynamics act concurrently, and the entanglement builds up during the pumping itself. The negativity is a property of the instantaneous two-qubit state, so this maximum is available as a resource as any later one, provided the pump laser acts on the mediator's dot alone. The following recurrences decay from $0.56$ to $\approx0.35$ over the next $\sim300\,$ns. Two effects contribute. The qubits' own dephasing reduces every recurrence, as the mediator, unlike the qubits, is the only subsystem reset by the protocol. In addition, the residual population of the mediator outside $m=-S$ (about $11\%$, mostly in $m=-S+1$) makes the qubit dynamics a weighted sum of oscillations with slightly different frequencies, one for each mediator sublevel; their beating produces the slight modulation of the late recurrences. A \emph{single} pulse thus suffices to purify the mediator on a timescale that is, for all practical purposes, permanent; the attainable entanglement is then limited by the residual mixing left by the backward pumping channel and by the qubits' dephasing.

\subsection{Thermal robustness of the first-peak negativity}
\label{sec:results-temperature}

\begin{figure}
\centering
\includegraphics[width=1.0\columnwidth]{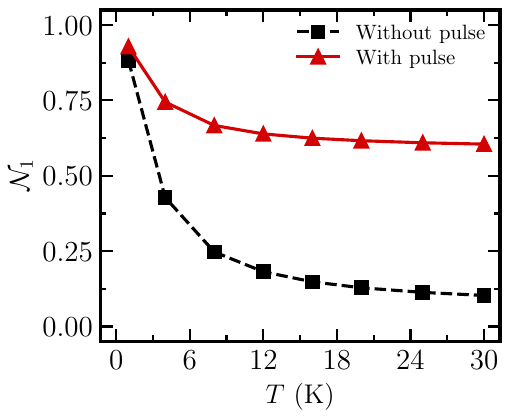}
\caption{First-recurrence negativity $\Neg_1$ as a function of temperature, with dephasing and mediator relaxation always active. Without the pulse (black squares), $\Neg_1$ falls from $\approx0.88$ at $T=1\,$K to $\approx0.10$ at $T=30\,$K. With a single purification pulse applied at $t=0$ (red triangles), $\Neg_1$ stays between $0.61$ and $0.67$ from $8$ to $30\,$K, and reaches $0.93$ at $1\,$K.}
\label{fig:NvsT}
\end{figure}

Finally, we return to the temperature dependence introduced in Sec.~\ref{sec:results-problem} and assess the extent to which a single purification pulse restores the entanglement lost to thermal mixing, across the full range explored in this work. Figure~\ref{fig:NvsT} compares $\Neg_1$ as a function of temperature, with and without the pulse ($\kappa_{\mathrm p\downarrow}=0.013\,\mu$eV, $\kappa_{\mathrm p\uparrow}=0.104\,\kappa_{\mathrm p\downarrow}$, $\taup=80\,$ns).

Without the pulse, $\Neg_1$ decreases by nearly an order of magnitude between $T=1\,$K and $T=30\,$K, consistent with the individual curves already shown in Fig.~\ref{fig:problem}. With the pulse applied, $\Neg_1$ becomes nearly independent of temperature above a few kelvin: it varies only from $0.67$ at $8\,$K to $0.61$ at $30\,$K, while the value without the pulse falls from $0.25$ to $0.10$ over the same range. The gain therefore grows with temperature, from a factor of $3.5$ at $12\,$K to $6$ at $30\,$K. At $1\,$K, where the thermal state is already strongly polarized, the pulse raises $\Neg_1$ from $0.88$ to $0.93$. Rather than relying on a low operating temperature to favor the mediator's extremal-state population, the pulse prepares that population on demand, up to the limit set by the pumping cycle itself. As established in Sec.~\ref{sec:results-mechanism}, this preparation persists on a timescale of hundreds of microseconds after a single application.

\section{Conclusions}
\label{sec:conclusions}

We have shown that entanglement between two spin qubits mediated by a thermally mixed, high-spin magnetic impurity cannot be restored by any amount of coherent control acting on the mediator alone: unitary evolution exactly preserves the purity of any state it acts on, so a static field, or indeed any resonant coherent drive of the mediator's own degree of freedom, leaves it pinned at its thermal value no matter how strong or precisely tuned. We verified this numerically for a resonant drive of the same magnitude as the dissipative channel. Genuine dissipation is not a design choice but a necessity, physically realized here through resonant optical pumping of the mediator's spin.

Having established that dissipation is necessary, we further showed that it is not sufficient to make it as strong as possible. 
Because the same channel that purifies the mediator is a genuine Lindblad process, it unavoidably leaks decoherence into the qubits through the exchange coupling that mediates their entanglement, so the resulting entanglement peaks at a well-defined, finite pump strength rather than growing monotonically with it.

We fixed the pumping rates by fitting the effective dissipators to a microscopic model of the optical pumping cycle of a single Mn spin, which reproduces the measured orientation times and limits the mediator's purity to $\approx0.8$ through backward processes within the same cycle. At this experimental rate ($\kappa_{\mathrm p\downarrow}=0.013\,\mu$eV), a single $80\,$ns pulse purifies the mediator to $\mathrm{Tr}[\rho_M^2]\approx0.80$, and the resulting polarization persists for hundreds of microseconds, three orders of magnitude longer than the entangling dynamics considered here. The first-recurrence negativity rises from $0.18$ to $0.64$ at $12\,$K and stays above $0.6$ up to $30\,$K. The optimum lies at a pump rate about one order of magnitude higher, where $\Neg_1$ would reach $\approx0.93$; faster optical pumping schemes are therefore a direct route to stronger mediated entanglement.
No repeated pulsing is required: within the present model, the residual mixing left by the backward pumping channel and the qubits' dephasing are the remaining limitations.

Throughout, we grounded the parameters entering our model in reported experimental and theoretical values for a single Mn spin in a CdTe quantum dot (Table~\ref{tab:parameters}), and in a microscopic description of the optical pumping cycle with parameters from Ref.~\cite{smolenski2018mechanisms}, so the regime explored here sits within reach of optical pumping schemes already demonstrated in the literature~\cite{legall2009optical,goryca2009optical}. More broadly, these results are a concrete instance of a general principle: dissipation, rather than being merely an obstacle to be suppressed, can be engineered as an active resource for entanglement generation~\cite{verstraete2009,krauter2011}.%, provided its strength is tuned to balance the purification it provides against the decoherence it inevitably introduces through the same coupling that makes it useful. 

\begin{acknowledgments}
E.~M.~F. acknowledges FAPEMIG for financial support. F.~M.~S. acknowledges financial support from the National Institute of Science and Technology for Applied Quantum Computing through CNPq, process No.~408884/2024-0.
\end{acknowledgments}

\appendix

\section{Single mode versus continuum: why the pump must be dissipative}
\label{app:jc}

Consider coupling the mediator to a bosonic field at frequency $\omega=\varepsilon_M$ (resonant with the mediator's own splitting) via a Jaynes--Cummings-type interaction. Two limits are relevant.

\emph{Single mode} (cavity-like, unitary):
\begin{equation}
H_{\mathrm{field}}^{(1)}=\omega\,a^\dagger a, \qquad V_1=\Omega\left(a^\dagger S_- + a\,S_+\right).
\label{eq:jc-single}
\end{equation}
The joint mediator--field system evolves unitarily under $H_0+V+H_{\mathrm{field}}^{(1)}+V_1$. Because this is a closed bipartite system, any excitation transferred to the field mode is returned to the mediator after a finite Poincar\'e recurrence time (vacuum Rabi oscillations); no Born--Markov approximation is applicable, and $\mathrm{Tr}[\rho_M^2]$ cannot be permanently increased. In the numerical comparison of Sec.~\ref{sec:results-theorem}, this single-mode drive is applied semiclassically, with $a\to$ a prescribed oscillating field, $H_{\mathrm{drive}}(t)=\Omega(S_+e^{-i\omega t}+S_-e^{+i\omega t})$, which is the interaction-picture form of Eq.~\eqref{eq:jc-single}; the conclusion is unaffected by this choice, since both descriptions are unitary.

\emph{Continuum of modes} (reservoir, dissipative):
\begin{equation}
H_{\mathrm{field}}^{(2)}=\sum_k\omega_k\,a_k^\dagger a_k, \qquad V_2=\sum_k g_k\left(a_k^\dagger S_- + a_k S_+\right).
\label{eq:jc-continuum}
\end{equation}
Tracing out the continuum under the standard Born--Markov approximation (weak coupling, spectral density smooth near $\omega_k\approx\varepsilon_M$) yields exactly the Lindblad dissipator $L_{\mathrm p\downarrow}$ of Eq.~\eqref{eq:pump},
\begin{equation}
\mathcal{D}[S_-]\rho = S_-\rho S_+ - \tfrac{1}{2}\{S_+S_-,\rho\},
\label{eq:golden-rule}
\end{equation}
\begin{equation}
\kappa_{\mathrm p\downarrow}=2\pi\,|g(\varepsilon_M)|^2\,\rho_{\mathrm{modes}}(\varepsilon_M),
\label{eq:golden-rule-rate}
\end{equation}
via the standard Wigner--Weisskopf/Fermi's Golden Rule argument. A reservoir with mean occupation $\bar N$ at $\omega\approx\varepsilon_M$ would add the reverse channel $\sqrt{\kappa_{\mathrm p\downarrow}\bar N/(\bar N+1)}\,S_+$, which has the form of $L_{\mathrm p\uparrow}$. Physically, $H_{\mathrm{field}}^{(2)}+V_2$ represents the continuum of electromagnetic vacuum modes into which a spontaneously emitted photon escapes irreversibly, the same mechanism underlying the optical pumping schemes of Refs.~\cite{legall2009optical,goryca2009optical}.

The distinction between Eqs.~\eqref{eq:jc-single} and \eqref{eq:jc-continuum} is not one of coupling strength but of reservoir dimensionality: a single mode always returns its excitation (bounded recurrence); a continuum carries it away for good. This irreversibility, not the strength of the coupling, is what enables purification, as verified numerically in Fig.~\ref{fig:theorem}.

\section{Time units: converting the $\hbar=1$ dynamics to nanoseconds}
\label{app:units}

Throughout this work we set $\hbar=1$ (Sec.~\ref{sec:theory}), so that every energy scale entering Eq.~\eqref{eq:lindblad}, namely the Hamiltonian $H$, the exchange couplings $J_1,J_2$, the Zeeman splittings $\varepsilon_M,\varepsilon_{\mathrm{QD}}$, and every dissipation rate ($\Gamma_\downarrow,\Gamma_\uparrow,\gamma_M,\gamma_Q,\kappa_{\mathrm p\downarrow},\kappa_{\mathrm p\uparrow}$), is expressed in a single unit, $\mu$eV (Table~\ref{tab:parameters}). Under this convention, the time variable $t$ appearing in Eq.~\eqref{eq:lindblad}, and hence the time grid supplied to the numerical integrator, is not itself a real time: restoring $\hbar$ explicitly, the master equation reads $d\rho/dt_{\mathrm{real}} = -i[H,\rho]/\hbar + \hbar^{-1}\sum_k\mathcal{D}[L_k]\rho$, so the dimensionless-looking $t$ used throughout our numerics is related to the physical elapsed time by
\begin{equation}
t_{\mathrm{real}} = \hbar\, t_{\mathrm{sim}},
\label{eq:hbar-conversion}
\end{equation}
with $\hbar = 6.582119569\times10^{-16}\,\mathrm{eV\,s} = 0.658212\,\mu\mathrm{eV\,ns}$. Equation~\eqref{eq:hbar-conversion} is the single conversion factor used to produce every time axis shown in Figs.~\ref{fig:problem}--\ref{fig:NvsT}: the raw output times of the simulation (in units of inverse $\mu$eV) are multiplied by this numerical value of $\hbar$ to obtain nanoseconds.

The same constant underlies the ``real-time equivalent'' column of Table~\ref{tab:parameters}, but appears there in two distinct guises, since it answers two different physical questions:

\begin{itemize}
\item[(i)] For a \emph{dissipation rate} $\Gamma$ (e.g.\ $\Gamma_\downarrow$, $\gamma_M$, $\gamma_Q$, $\kappa_{\mathrm p\downarrow}$), the relevant real-time quantity is the associated exponential lifetime,
\begin{equation}
\tau_{1/e} = \hbar/\Gamma,
\label{eq:lifetime-conversion}
\end{equation}
exactly as it enters Eq.~\eqref{eq:lindblad}, with no factor of $2\pi$. For instance, $\kappa_{\mathrm p\downarrow}=0.013\,\mu$eV corresponds to $\tau_{1/e}=51\,$ns, and $\Gamma_\downarrow=1.65\times10^{-6}\,\mu$eV to $\tau_{1/e}=400\,\mu$s, both quoted in Table~\ref{tab:parameters}.
\item[(ii)] For a \emph{coherent energy splitting or coupling} ($\varepsilon_M$, $\varepsilon_{\mathrm{QD}}$, $J_1$, $J_2$), the more physically meaningful real-time quantity is instead an oscillation period (or frequency), related to $E$ through the ordinary Planck constant $h=2\pi\hbar$,
\begin{equation}
T_{\mathrm{osc}} = h/E, \qquad \nu = E/h,
\label{eq:period-conversion}
\end{equation}
the standard spectroscopy convention. For instance, $J_1=J_2=1\,\mu$eV corresponds to a $4.14\,$ns oscillation period ($0.242\,$GHz), as quoted in Table~\ref{tab:parameters}.
\end{itemize}

The two conversions share the same underlying constant $\hbar$ but package it differently ($\hbar$ alone versus $h=2\pi\hbar$) because they answer different questions: how long an exponential decay takes, versus what frequency a coherent oscillation corresponds to. 
This difference reflects the nature of each question, not a different value of $\hbar$. Equation~\eqref{eq:hbar-conversion} is simply the special case of Eq.~\eqref{eq:lifetime-conversion} in which the quantity being converted is already an elapsed simulation time rather than a decay rate.

\bibliography{references_ellen}

\end{document}